\documentclass{article}
\usepackage{lmodern}
\usepackage[margin=1in]{geometry}
\usepackage{amsmath,amssymb,amsfonts,amsthm,mathtools}
\usepackage{microtype}
\usepackage{graphicx}
\usepackage{multirow,booktabs,enumitem,subcaption,float}
\usepackage{setspace}
\usepackage{xcolor}
\usepackage{tikz}
\usetikzlibrary{arrows.meta,positioning,calc}
\usepackage[backend=biber,style=authoryear,sorting=nyt,useprefix=true, uniquename=false,maxcitenames=2,maxbibnames=99,doi=false,url=false,isbn=false]{biblatex}
\usepackage[colorlinks=true,linkcolor=blue!45!black,citecolor=blue!45!black,urlcolor=blue!45!black]{hyperref}

\numberwithin{equation}{section}

\newtheorem{theorem}{Theorem}[section]
\newtheorem{definition}[theorem]{Definition}
\newtheorem{proposition}[theorem]{Proposition}
\newtheorem{lemma}[theorem]{Lemma}
\newtheorem{remark}[theorem]{Remark}

\newcommand{\R}{\mathbb{R}}
\newcommand{\E}{\mathbb{E}}
\newcommand{\Prob}{\mathbb{P}}

\newcommand{\cE}{\mathcal{E}}
\newcommand{\cI}{\mathcal{I}}
\newcommand{\cW}{\mathcal{W}}
\newcommand{\cR}{\mathcal{R}}
\newcommand{\cP}{\mathcal{P}}

\title{Sequential Monitoring of Image Streams Using Energy Statistics}
\author{Shreetama Bhuniya$^{1}$\thanks{These authors contributed equally.}\phantom{$^*$}, Sourav De$^{1}$\footnotemark[1]\phantom{$^*$}, Anik Roy$^{2}$, Partha Sarathi Mukherjee$^{1}$\\
$^{1}$Indian Statistical Institute, Kolkata\\
$^{2}$Emory University, Atlanta, USA}
\date{\today}

\begin{document}

\maketitle

\begin{abstract}

Sequential monitoring of image data is a high-velocity, high-dimensional big data problem in statistics and machine learning, with wide applications in manufacturing industries, satellite imaging, medical diagnostics, and remote sensing. This is a challenging task because image intensity functions are often contaminated by noise and contain discontinuities and other complicated spatial structures. In the literature, a vast majority of image-monitoring methods rely on changes in image intensity values to detect out-of-control images. However, such approaches are often unreasonable in applications where insignificant changes in image contrast or background intensity should not indicate an out-of-control image as long as the boundaries of the image objects remain unchanged. In this article, we propose a nonparametric Phase II image-monitoring procedure based on energy distance that directly monitors changes in the distribution of image boundaries. The proposed method constructs a moving-window energy-distance-based control statistic by comparing the distribution of recently observed image boundaries with that of a fixed in-control reference sample. The central idea behind the proposed boundary-based approach is twofold: first, to reduce the dimensionality of high-resolution images by retaining relevant edge information, and second, to provide a robust monitoring procedure that is less sensitive to insignificant smooth changes in image intensity. The proposed construction offers a computationally efficient distributional approach to image monitoring without requiring a parametric model for out-of-control images. Theoretical justifications and superior performance in numerical comparisons ensure that the proposed control chart can be useful to many researchers and practitioners.

\end{abstract}

\section{Introduction}\label{intro}
\begin{figure}[H]
\centering
\includegraphics[width=0.85\textwidth]{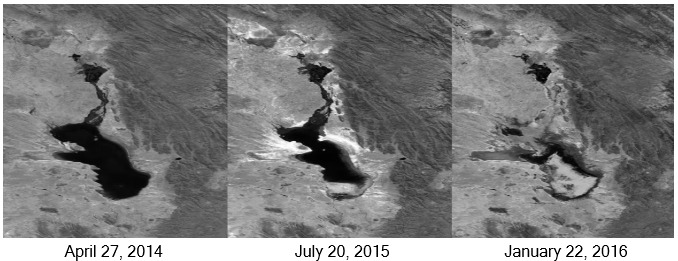}   
\caption{Satellite images of Lake Poop\'o in Bolivia. (Source: European Space Agency's (ESA) mini-satellite Proba-V.)}
\label{fig:poopo_lake}
\end{figure}

Recent advances in image acquisition technologies have facilitated the routine collection of massive volumes of image data across a wide range of scientific disciplines. Such images play a crucial role in monitoring and surveillance applications in diverse fields, including manufacturing, medicine, and environmental and ecological sciences. In manufacturing industries, the monitoring of image sequences has significant applications in quality control and inspection processes. In the medical field, advanced imaging modalities, such as X-rays, computed tomography (CT), magnetic resonance imaging (MRI), and functional MRI (fMRI), are extensively employed for diagnostic purposes and disease monitoring. For example, brain MRI scans are commonly used to assess the progression of brain tumors over time. Similarly, in satellite imaging and remote sensing, monitoring changes in the Earth's surface constitutes an important and challenging research problem. Figure \ref{fig:poopo_lake} presents a sequence of satellite images of Lake Poopo acquired by the European Space Agency's (ESA) mini-satellite Proba-V. A comparison of these images clearly reveals a substantial reduction in the lake's surface area over time, primarily attributable to the rapid evaporation of its water. This phenomenon highlights the potential impact of climate change on the region. Beyond monitoring changes in water bodies, satellite imaging and remote sensing have extensive applications in several other domains, including agriculture, forestry, ecology and ecosystem studies, coastal resource management, and environmental monitoring, among others.

Image surveillance or monitoring is, therefore, a fundamental research problem with numerous applications in various fields of science. The pivotal step for the surveillance of a sequence of images is \textit{image comparison}, i.e., constructing a statistical testing procedure to check whether a significant difference between two images captured at different time points exists. There are two major problems in image monitoring. The first major problem is the high dimensionality, as each image contains several thousand pixels. The second problem is geometric misalignment of various images. This paper addresses the first problem by reducing dimensionality through edge-pixel detection and then applying energy statistics to construct an image monitoring procedure.

\subsection{Related literature}\label{literature}
Image comparison and monitoring have attracted considerable attention in recent years owing to their wide range of applications. However, despite this growing interest, the problem of image comparison in the presence of geometric misalignment has received relatively limited attention. Early developments in nonparametric regression primarily focused on the comparison of regression curves \parencite{king1991testing, kulasekera1995comparison, wang2017comparison}. The underlying ideas for nonparametric comparison of image surfaces are closely related to nonparametric analysis of covariance (ANCOVA) involving multiple covariates. In this context, \textcite{bowman2006comparing} proposed a generalized ANOVA-type test for comparing regression surfaces, whereas \textcite{wang2010nonparametric} introduced a more flexible approach based on the $\mathcal{L}^2$ distance between estimated surfaces. Both approaches extend the framework developed by \textcite{dette2001nonparametric}. More recently, \textcite{zhao2020comparison} developed semiparametric testing procedures for comparing nonlinear curves and surfaces.

Statistical process control (SPC) charts \parencite{qiu2013introduction} were initially developed for monitoring univariate processes. Subsequently, considerable research has extended these methods to multivariate processes, profile monitoring, and image-based monitoring. In modern manufacturing, images have become a common form of data for quality inspection and control. \textcite{megahed2011review} provided a comprehensive review of image-based process monitoring methods. Many existing approaches begin by identifying prespecified regions of interest (ROIs) within each image and subsequently construct control charts using summary statistics derived from these regions, such as average intensity over time. Examples include the generalized likelihood ratio (GLR) chart \parencite{megahed2012spatiotemporal} and the multivariate GLR (MGLR) charts \parencite{he2016image,okhrin2020new}. Other approaches formulate image monitoring as a two-dimensional profile monitoring problem and employ feature-based schemes involving wavelet coefficients observed over time \parencite{Koosha2017}. In a related context, \textcite{yan2017anomaly, kang2022statistical} considered anomaly detection in smooth image intensity surfaces for quality inspection.

In computer vision and remote sensing, considerable research has also been devoted to multi-temporal image comparison \parencite{de2019unsupervised, qu2021change}. Furthermore, deep learning techniques have demonstrated impressive performance in image change detection \parencite{fang2021snunet}. The existing literature in this area can broadly be classified into supervised and unsupervised approaches. Supervised methods rely on ground-truth information \parencite{chen2021adversarial}, whereas unsupervised methods aim to identify changes by comparing images of the same scene acquired at different time points without requiring additional labeled information \parencite{wu2021unsupervised}. More recently, research has increasingly focused on large-scale monitoring of tensor-valued data \parencite{li2024dynamic}. Since digital images can naturally be represented as second-order tensors, these approaches are potentially well suited to image monitoring. Nevertheless, most existing methodologies rely on raw pixel intensities or parametric models for out-of-control images, making them sensitive to nuisance illumination or contrast shifts that leave object boundaries intact.

\subsection{Objectives and contributions of this paper}

This paper develops a Phase II monitoring procedure for geometrically aligned image streams in which changes in object boundaries are of primary interest. The main contributions are as follows.
\begin{enumerate}
    \item \textbf{Boundary-based distributional monitoring.} We use energy distance to compare the distribution of recent binary edge representations with that of a fixed reference sample. The chart requires no model for out-of-control images and is unaffected by intensity variations that leave the detected boundaries unchanged.

    \item \textbf{Exact reduction to relevant pixels.}
    We construct a common coordinate set containing the edge supports of both samples (enlarged by a $3\times 3$ neighborhood) by retaining only pixels near a detected edge in either sample, preserving all pairwise Euclidean distances and hence the energy statistic exactly.
    
    \item \textbf{Sequential control-limit calibration.}
    We develop a parametric bootstrap procedure to select a control limit targeting a prespecified in-control average run length. The calibration retains the observed reference sample and reproduces pre-processing and overlapping moving windows used during monitoring, accounting for the resulting dependence between successive charting statistics.

\end{enumerate}

The remainder of this paper is organized as follows. Section~\ref{sec:method} presents the proposed image monitoring methods using dimensionality reduction and energy statistics. Related theoretical properties are discussed in Section~\ref{sec:theory}. We present numerical simulations and a real data analysis in Section~\ref{sec:num}. Section~\ref{sec:conclusion} provides concluding remarks.

\section{Proposed Methodology}\label{sec:method}
In this article, we primarily focus on Phase II monitoring of sequential image data, aiming to detect significant departures from the in-control boundary structure. In Phase I, a reference sample is used to characterize the in-control process, estimate the monitoring parameters, and calibrate the control limit for a desired in-control average run length (ARL). In Phase II, newly observed images are monitored sequentially to detect potential changes as quickly as possible. The proposed monitoring procedure consists of the following major steps.

\subsection{Model description}\label{sec:setup}
Suppose that a sequence of $n\times n$ grayscale images $\{Z_t\}_{t\ge -w+1}$ is observed from an imaging process, where $w\ge2$ denotes a prespecified window size. Let $\cP=\{1,\ldots,n\}^2$ denote the pixel lattice. Under the jump regression analysis (JRA)
framework \parencite{qiu2005image}, a
two-dimensional grayscale image can be regarded as a noisy
observation of a potentially discontinuous intensity surface. We model the intensity observed at pixel $(i,j)\in\cP$ and time $t$ as
\begin{equation}\label{eq:image}
 Z_t(i,j)=f_t(i,j)+\varepsilon_t(i,j), \qquad (i,j)\in\cP,
\end{equation}
where $f_t$ is the underlying image intensity surface and $\varepsilon_t$ represents measurement noise. In our setting, the noise variables are independent across time, have mean zero, and common variance $\sigma^2$. 
We assume that $f_t$ is
piecewise continuous, with a finite number of continuity
regions, and may have abrupt intensity changes along
the boundaries of image objects. The locations of these
discontinuities are generally curves, commonly referred
to as jump location curves (JLCs) in the JRA literature.
These curves represent the boundaries of image objects
and constitute the primary features of interest in our
monitoring procedure. For detecting a step shift in image data, we consider the change-point model
\begin{equation}\label{eq:step}
f_t=
\begin{cases}
f^{(0)}, & t \le \tau,\\
f^{(1)}, & t > \tau,
\end{cases}
\end{equation}
where $\tau$ denotes the unknown change point (the last in-control time point), and $f^{(0)}$ and $f^{(1)}$ represent the underlying intensity surfaces corresponding to the in-control and out-of-control processes, respectively. Although \eqref{eq:step} represents a canonical step shift, the proposed nonparametric chart requires no parametric model for out-of-control images and naturally detects gradual or multi-stage boundary drifts, as evaluated in Section~\ref{sec:num}. To characterize the in-control boundary structure in Phase II, we assume that a reference sample of $w$ in-control images, $Z_{-w+1},\ldots,Z_0$, is available from Phase I. This reference sample serves as the basis for detecting changes in the boundaries of subsequently observed images.

\subsection{Image pre-processing and edge representation}\label{sec:edge}
The first step in image monitoring is to ensure that the observed images are comparable over time. When necessary, images are converted to grayscale, resized to a common resolution, and standardized using fixed pre-processing settings. Geometric misalignment is another important issue, which may arise due to changes in the relative positions of the camera and image objects. Several image registration techniques have been developed in the literature to address this problem \parencite{xing2011intensity,feng2018difference,das2024image}. Throughout this paper, we assume that the observed images are geometrically aligned and that the pre-processing settings remain unchanged during Phase II monitoring.

\paragraph{Estimation of jump location curves} Since the proposed monitoring procedure is based on the distribution of edge pixels, identifying the boundaries or jump location curves (JLCs) of image objects is an essential step. Although several edge detection methods are available in the literature \parencite{sun2007jump,xie2015holistically,ronneberger2015u}, we employ the classical Canny edge detection algorithm \parencite{canny1986computational} for its simplicity and effectiveness in detecting image boundaries. The procedure begins by smoothing the observed image $Z_t$ using a Gaussian filter $G_b$ with bandwidth $b>0$, followed by computing the horizontal and vertical gradients using Sobel operators $S_x$ and $S_y$, respectively:
\begin{equation}\label{eq:grad}
H_{x,t}=S_x*(G_b*Z_t),\qquad
H_{y,t}=S_y*(G_b*Z_t),
\end{equation}
where $*$ denotes discrete convolution. The gradient magnitude and direction are then calculated as
\begin{equation}\label{eq:gradmag}
G_t=\sqrt{H_{x,t}^2+H_{y,t}^2},\qquad
\theta_t=\operatorname{atan2}(H_{y,t},H_{x,t}).
\end{equation}
Next, non-maximum suppression is applied to retain pixels with locally maximum gradient magnitudes along the gradient direction, with the objective of obtaining thin and well-defined edges. This is followed by double thresholding using two prespecified thresholds $u_L<u_H$ to classify the pixels into strong edges, weak edges, and non-edges. Finally, edge tracking by hysteresis is performed to retain weak edge pixels connected to strong edges and eliminate isolated weak edges that may arise due to noise.

In the monitoring context, let $\mathcal C_{\vartheta}$ denote the overall pre-processing procedure, including any fixed grayscale standardization and Canny edge detection, where $\vartheta$ represents the preprocessed parameters, including the smoothing bandwidth $b$ and thresholds $(u_L,u_H)$. These parameters are selected using Phase I images and kept fixed throughout Phase II monitoring. The resulting binary edge matrix for each observed image $Z_t$ is defined as
\begin{equation}\label{eq:canny-transform}
D_t=\mathcal C_{\vartheta}(Z_t)\in\{0,1\}^{n\times n},
\end{equation}
where $D_t(i,j)=1$ indicates a detected edge pixel and $D_t(i,j)=0$ otherwise. Let $P_0$ denote the in-control distribution of $\operatorname{vec}(D_t)$. The primary objective of the proposed monitoring procedure is to detect significant departures from $P_0$ as new images are observed sequentially.

\subsection{Online monitoring}\label{sec:online_monitoring}

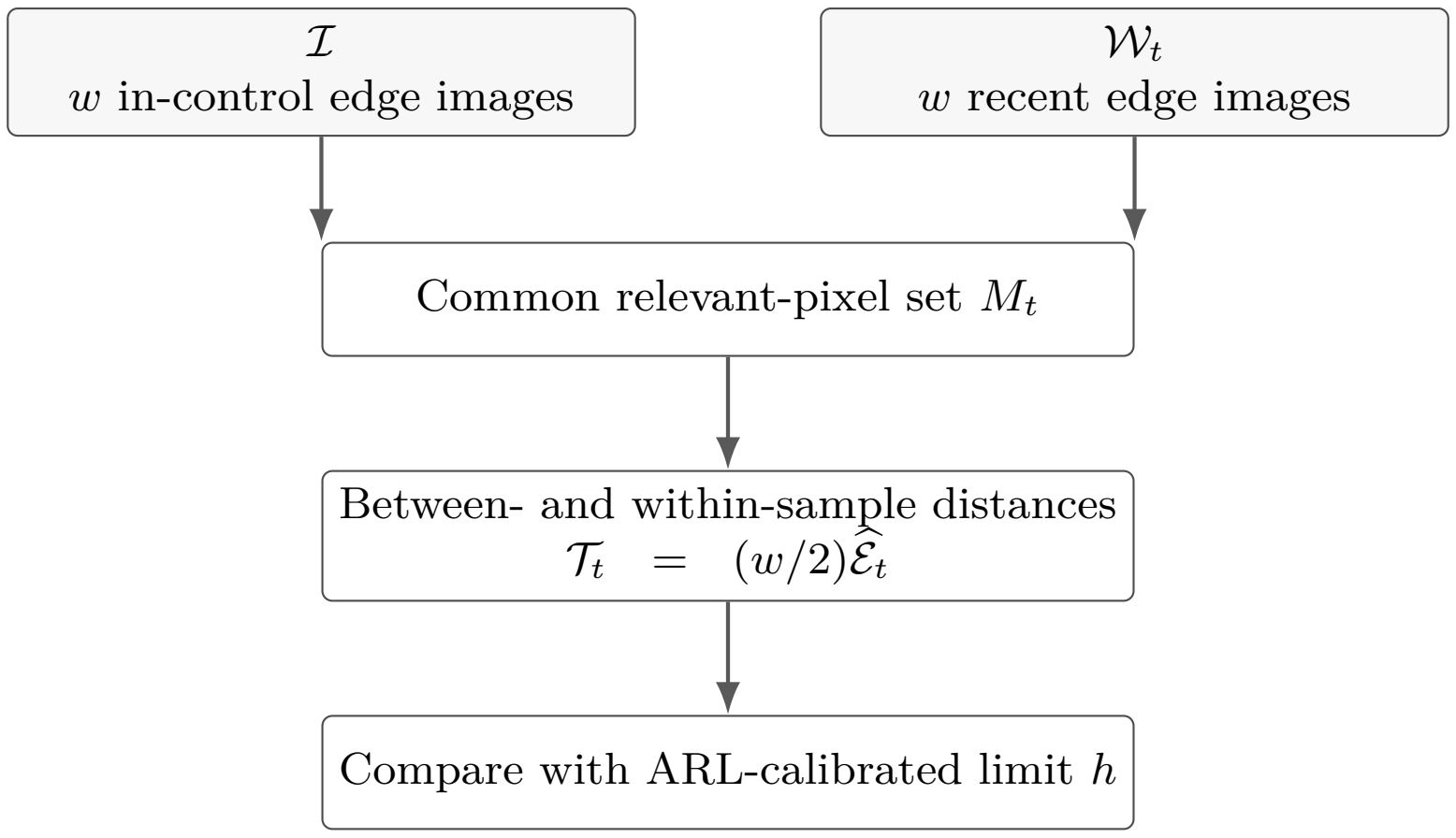
\begin{figure}[H]
\centering
\begin{tikzpicture}[
  node distance=8mm and 13mm,
  box/.style={draw=black!70, rounded corners=2pt, align=center,
              minimum height=9mm, text width=42mm, font=\small, fill=black!3},
  stagebox/.style={draw=black!70, rounded corners=2pt, align=center,
               minimum height=8mm, text width=55mm, font=\small},
  flow/.style={-Latex,thick,draw=black!65}]
  \node[box] (ic) {$\mathcal I$\\ $w$ in-control edge images};
  \node[box,right=13mm of ic] (mon) {$\mathcal W_t$\\ $w$ recent edge images};
  \node[stagebox,below=12mm of $(ic)!0.5!(mon)$] (support) {Common relevant-pixel set $M_t$};
  \node[stagebox,below=8mm of support] (energy) {Between- and within-sample distances\\ $\mathcal T_t=(w/2)\widehat{\mathcal E}_t$};
  \node[stagebox,below=8mm of energy] (limit) {Compare with ARL-calibrated limit $h$};
  \draw[flow] (ic.south) -- (support.north west);
  \draw[flow] (mon.south) -- (support.north east);
  \draw[flow] (support) -- (energy);
  \draw[flow] (energy) -- (limit);
\end{tikzpicture}
\caption{Schematic of the Phase II energy-distance image-monitoring procedure. The reference set stays fixed, whereas the moving sample is updated whenever a new image arrives.}
\label{fig:scheme}
\end{figure}

Construct the fixed reference set from the $w$ Phase I images as
$ \cI=\{D_{-w+1},D_{-w+2},\ldots,D_0\}.$ After the process enters Phase II, the most recent $w$ preprocessed images at monitoring time $t$ form the moving window $\cW_t=\{D_{t-w+1},D_{t-w+2},\ldots,D_t\},\, t\ge w$. Note that the reference set remains fixed, while $\cW_t$ is updated by removing $D_{t-w}$ and adding $D_t$ whenever a new image arrives. We begin formal charting at $t=w$, when the first complete Phase II window is available. The monitoring question at time $t$ is whether the recent edge maps in $\cW_t$ appear compatible with those in the reference sample $\cI$. Under stable in-control operation, the two sets have the same marginal distribution, although consecutive moving windows overlap and their charting statistics will be strongly dependent over time. Following a sustained process change, the moving window gradually replaces earlier in-control images with out-of-control images. Figure~\ref{fig:scheme} illustrates the resulting image-monitoring construction.

\subsubsection{Reduction to relevant edge pixels}\label{sec:reduction}
Every binary image $D_t$ contains $n^2$ coordinates, many of which are zero because no object boundary is present. To avoid repeatedly processing large background regions, we select a common set of relevant pixels for the reference and current moving samples. Let $ S_t=\bigcup_{D\in\cI\cup\cW_t}\big\{(i,j)\in\cP:D(i,j)=1\big\}$ be the union of the observed edge supports. We enlarge this set by a $3\times3$ neighborhood:
\begin{equation}\label{eq:Mt}
 M_t=\left\{(i,j)\in\cP:\;\exists\,(k,l)\in\{-1,0,1\}^2\ \text{such that}\ (i+k,j+l)\in S_t\right\}.
\end{equation}
The intersection with $\cP$ is implicit in \eqref{eq:Mt}, ensuring that pixels outside image boundaries are not used. Write $M_t=\{\alpha_1,\ldots,\alpha_{r_t}\}$ in lexicographical order, where $r_t=|M_t|$. For any binary image $D$ appearing in either sample, define its reduced vector by $ \boldsymbol D^{(M_t)}=\big(D(\alpha_1),\ldots,D(\alpha_{r_t})\big)^\top\in\R^{r_t}$. Thus, the two windows become
\begin{align*}
 \cI^{(M_t)}&=\{\boldsymbol D_{-w+1}^{(M_t)},\ldots,\boldsymbol D_0^{(M_t)}\},\\
 \cW_t^{(M_t)}&=\{\boldsymbol D_{t-w+1}^{(M_t)},\ldots,\boldsymbol D_t^{(M_t)}\}.
\end{align*}
Crucially, a \emph{single} $M_t$ is used for every member of both samples in a given comparison. Computing distances after separately selecting the nonzero coordinates of each image would be invalid because those reduced vectors would not share the same coordinate system. Note that the enlarged set $M_t$ is convenient for retaining a small neighborhood of each detected boundary, but that enlargement is not required for the distance identity proved in Proposition~\ref{prop:dim_reduction}. Since $S_t\subseteq M_t$, every omitted coordinate equals zero for all $2w$ images. Hence all Euclidean distances, and therefore the energy statistic, are exactly the same whether evaluated from the full binary maps or their reduced representations. 

\subsubsection{An energy statistic for monitoring image boundaries}\label{sec:energy}

Energy distance is a distributional discrepancy constructed from expected between-sample and within-sample distances \parencite{szekely2013}. For distributions $P$ and $Q$ on $\R^{n^2}$ with finite first moments, write
\begin{equation*}
 \cE(P,Q)=2\E\|X-Y\|_2-\E\|X-X'\|_2-\E\|Y-Y'\|_2,
\end{equation*}
where $X,X'$ are independent draws from $P$ and $Y,Y'$ are independent draws from $Q$. Euclidean energy distance is nonnegative and equals zero precisely when $P=Q$. Consequently, a nonzero value indicates a difference in the distribution of the vectorized edge images, rather than necessarily a difference in their mean images. At a monitoring time $t\ge w$, we estimate the energy distance using the following three quantities, all computed after the common reduction in Section~\ref{sec:reduction}:
\begin{align}
\cR\big(\cI^{(M_t)},\cW_t^{(M_t)}\big)
&=\frac{1}{w^2}\sum_{i=-w+1}^{0}\sum_{j=t-w+1}^{t}
 \big\|\boldsymbol D_i^{(M_t)}-\boldsymbol D_j^{(M_t)}\big\|_2,\label{eq:Rbetween}\\
\cR\big(\cI^{(M_t)},\cI^{(M_t)}\big)
&=\frac{1}{w^2}\sum_{i=-w+1}^{0}\sum_{j=-w+1}^{0}
 \big\|\boldsymbol D_i^{(M_t)}-\boldsymbol D_j^{(M_t)}\big\|_2,\label{eq:Rref}\\
\cR\big(\cW_t^{(M_t)},\cW_t^{(M_t)}\big)
&=\frac{1}{w^2}\sum_{i=t-w+1}^{t}\sum_{j=t-w+1}^{t}
 \big\|\boldsymbol D_i^{(M_t)}-\boldsymbol D_j^{(M_t)}\big\|_2.\label{eq:Rwindow}
\end{align}
The first term averages the distances between the in-control reference and current images. The second and third terms measure the natural variation within each sample. The resulting empirical energy distance is
\begin{equation}\label{eq:Ehat}
 \widehat{\cE}_t=2\cR\big(\cI^{(M_t)},\cW_t^{(M_t)}\big)
 -\cR\big(\cI^{(M_t)},\cI^{(M_t)}\big)
 -\cR\big(\cW_t^{(M_t)},\cW_t^{(M_t)}\big).
\end{equation}
Both within-sample terms are evaluated with denominators $w^2$, including zero diagonal distances. Therefore, \eqref{eq:Ehat} is the conventional nonnegative, \emph{biased} two-sample $V$-statistic estimator of energy distance, not its unbiased $U$-statistic counterpart. We retain this version because of non-negativity, and the corresponding standard two-sample scaling is convenient for chart construction.

We define the proposed global charting statistic by
\begin{equation}\label{eq:chartstat}
 {\mathcal T_t=\frac{w}{2}\widehat{\cE}_t,\qquad t\ge w.}
\end{equation}
Large values indicate that the recent boundary-image sample is relatively dissimilar to the in-control reference after accounting for variation within the two sets. The equality of the two underlying distributions is the relevant null hypothesis for a single complete window:
\begin{equation}\label{eq:null}
 H_{0,t}:\ P_t=P_0
 \qquad\text{versus}\qquad
 H_{1,t}:\ P_t\ne P_0,
\end{equation}
where $P_t$ denotes the marginal distribution represented by the observations within the current window; immediately after an abrupt change, this may be a mixture of pre- and post-change distributions. We use \eqref{eq:null} to explain the statistic, but the actual online decision is made through a control chart, not through a sequence of separately calibrated level-$\alpha$ hypothesis tests.

For a control limit $h>0$, define the stopping time and monitoring run length by
\begin{equation}\label{eq:stopping}
 \mathsf{S}(h)=\inf\{t\ge w:\mathcal T_t>h\},\qquad
 \mathrm{RL}(h)=\mathsf S(h)-w+1,
\end{equation}
with the convention $\inf\varnothing=\infty$. The chart declares an out-of-control signal at the first $t$ satisfying $\mathcal T_t>h$. The run length counts charting decisions starting from the first complete Phase II window; the actual number of Phase II images acquired by the alarm is $\mathrm{RL}(h)+w-1$.

\subsection{Determination of the control limit with prefixed $\mathrm{ARL}_0$}
\label{sec:control_limit}

The control limit $h$, defined in \eqref{eq:stopping}, is determined using a parametric bootstrap procedure to achieve a prespecified in-control average run length ($\mathrm{ARL}_0$). We calibrate $h$ by simulating the sequential monitoring procedure under the in-control condition. Let $Z_{-w+1},\ldots,Z_0$ denote the Phase I reference images. Following the proposed bootstrap construction, we first obtain a denoised image $Y_k$ by applying a $3\times 3$ median filter multiple times to $Z_k$, for $k=-w+1,\ldots,0$. The corresponding residual image is
defined by $\widehat{\varepsilon}_k=Z_k-Y_k.$ We estimate the noise variance $\sigma^2$ by the sample
variance $\widehat{\sigma}^2$ of the pooled residuals
$\{\widehat{\varepsilon}_k(i,j):
k=-w+1,\ldots,0;\ i,j=1,\ldots,n\}$. To generate a synthetic in-control image at monitoring time
$s$, we randomly select an index $L_s$ with replacement
from $\{-w+1,\ldots,0\}$ and construct $Z_s^*=Y_{L_s}+\eta_s,$ where $L_s\overset{\mathrm{iid}}{\sim}
\operatorname{Unif}\{-w+1,\ldots,0\}$, and  $\eta_s(i,j)\overset{\mathrm{iid}}{\sim}
N(0,\lambda^2\widehat{\sigma}^2)$. Here, $\lambda>0$ is a prespecified tuning parameter controlling the noise level of the simulated images, compensating for any bias in $\widehat{\sigma}^2$ introduced by median-filter smoothing near sharp boundaries or periodic textures. Thus,
the bootstrap procedure preserves the estimated in-control
image structures while introducing random measurement
variation. Applying the same pre-processing operation $\mathcal{C}_{\vartheta}$ (including edge detection) to each simulated image yields $D_s^*=\mathcal{C}_{\vartheta}(Z_s^*)$. Now, given fixed  Phase I reference samples $\mathcal I=
\{\mathcal C_{\vartheta}(Z_{-w+1}),\ldots,
\mathcal C_{\vartheta}(Z_0)\}$, for $t\geq w$, the bootstrap generated 
moving-window sample is constructed as $\mathcal W_t^*=
\{D_{t-w+1}^*,\ldots,D_t^*\}.$ Using $\mathcal I$ and $\mathcal W_t^*$, we apply the
same dimension-reduction procedure and compute the
corresponding energy-distance-based charting statistic
$\mathcal T_t^*$. For a given control limit $h$, synthetic in-control
images are generated sequentially until the monitoring
statistic exceeds $h$. The corresponding run length is
defined as
\begin{equation*}
\mathrm{RL}^*(h) = \inf\{t\ge w:\mathcal{T}_t^*>h\}-w+1,
\end{equation*}
which counts the number of evaluated monitoring windows
until the first signal. Repeating the above procedure independently $B$ times,
we obtain the bootstrap run lengths
$\mathrm{RL}_1^*(h),\ldots,\mathrm{RL}_B^*(h)$. The in-control average
run length is then estimated by
\begin{equation*}
\widehat{\mathrm{ARL}}_0(h) = \frac{1}{B}\sum_{b=1}^{B}\mathrm{RL}_b^*(h).
\end{equation*}
Finally, the control limit $h$ is selected such that
$\widehat{\mathrm{ARL}}_0(h)$ is sufficiently close to the
prespecified target $\mathrm{ARL}_0$. Specifically, if the
estimated $\mathrm{ARL}_0$ is smaller than the target value,
the control limit is increased; otherwise, it is
decreased. A bisection algorithm can be employed to
identify the appropriate control limit. For a related
discussion of the searching algorithm, readers are
referred to \textcite{qiu2013introduction}. Note that throughout the bootstrap calibration, the reference
sample and pre-processing parameters are kept fixed,
and each synthetic image is processed in exactly the
same manner as an incoming Phase II image. 

\section{Theoretical Properties}\label{sec:theory}

\subsection{Energy Statistic}

\begin{definition}[Energy Distance]
Let $X$ and $Y$ be independent random vectors in $\R^d$ with $\E\|X\|_2 < \infty$ and $\E\|Y\|_2 < \infty$, and let $X'$ and $Y'$ be independent copies of $X$ and $Y$, respectively. The energy distance between the distributions $P_X$ and $P_Y$ of $X$ and $Y$ is defined \parencite{Szekely2004} as
\[
\cE(P_X, P_Y) = 2\,\E\|X - Y\|_2 - \E\|X - X'\|_2 - \E\|Y - Y'\|_2.
\]
\end{definition}

\begin{remark}
A fundamental property of the Euclidean energy distance is that $\cE(P_X, P_Y) \ge 0$, with $\cE(P_X, P_Y) = 0$ if and only if $P_X = P_Y$ \parencite{Szekely2004}.
\end{remark}

\begin{definition}[Sample Energy Statistic]
Let $X_1, \dots, X_{n_1} \overset{\mathrm{iid}}{\sim} P_0$ and $Y_1, \dots, Y_{n_2} \overset{\mathrm{iid}}{\sim} P_1$ be independent random samples in $\R^d$. Define the distance kernel
\[
\kappa(x_1, x_2, y_1, y_2) = \|x_1 - y_1\|_2 + \|x_2 - y_2\|_2 - \|x_1 - x_2\|_2 - \|y_1 - y_2\|_2.
\]
The empirical energy distance is the two-sample $V$-statistic \parencite{Szekely2004}
\[
\cE_{n_1,n_2} = \frac{1}{n_1^2 n_2^2} \sum_{i=1}^{n_1}\sum_{j=1}^{n_1}\sum_{l=1}^{n_2}\sum_{m=1}^{n_2} \kappa(X_i, X_j, Y_l, Y_m),
\]
and the associated two-sample energy test statistic is
\[
T_{n_1,n_2} = \frac{n_1 n_2}{n_1 + n_2}\cE_{n_1,n_2}.
\]
In particular, the online charting statistic $\mathcal{T}_t = \frac{w}{2}\widehat{\cE}_t$ in \eqref{eq:chartstat} corresponds to $T_{n_1,n_2}$ with $n_1 = n_2 = w$ applied to the vectorized binary edge maps in $\cI$ and $\cW_t$ (with $d = n^2$).
\end{definition}

\subsection{Two-Sample Hypothesis Test}

We first verify that the coordinate reduction in Section~\ref{sec:reduction} preserves the energy statistic identically.

\begin{proposition}[Exact Distance Preservation under Dimension Reduction]
\label{prop:dim_reduction}
Let $M_t$ be the common relevant-pixel set defined in \eqref{eq:Mt}. The empirical energy distance computed on the reduced sets $\cI^{(M_t)}$ and $\cW_t^{(M_t)}$ equals that computed on the full binary edge maps $\cI$ and $\cW_t$:
\[
\widehat{\cE}\big(\cI^{(M_t)}, \cW_t^{(M_t)}\big) = \widehat{\cE}(\cI, \cW_t).
\]
\end{proposition}

\begin{proof}
By construction of $M_t$, any binary edge map $D \in \cI \cup \cW_t$ satisfies $D(i,j) = 0$ for all $(i,j) \notin M_t$. Hence, for any $D, D' \in \cI \cup \cW_t$,
\[
\|D - D'\|_2
= \sqrt{\sum_{(i,j)\in\cP} (D(i,j) - D'(i,j))^2}
= \sqrt{\sum_{(i,j)\in M_t} (D(i,j) - D'(i,j))^2}
= \big\|\boldsymbol{D}^{(M_t)} - \boldsymbol{D}'^{(M_t)}\big\|_2.
\]
Because all pairwise Euclidean distances are preserved exactly, the resulting empirical energy distances and charting statistics are identical.
\end{proof}

Next, we show that a genuine change in the true underlying edge structure yields a strictly positive population energy distance despite edge-detection errors. Let $g_0, g_1 \in \{0,1\}^{n\times n}$ denote the deterministic true edge matrices corresponding to the intensity surfaces $f^{(0)}$ and $f^{(1)}$, and let $P_0$ and $P_1$ denote the distributions of the corresponding observed edge maps $\operatorname{vec}(D_0)$ and $\operatorname{vec}(D_1)$. Under each regime $k \in \{0,1\}$, define the pixelwise false-positive and false-negative edge probabilities by $p_{0,k}(i,j) = \Prob(D_k(i,j)=1)$ whenever $g_k(i,j)=0$, and $p_{1,k}(i,j) = \Prob(D_k(i,j)=0)$ whenever $g_k(i,j)=1$.

\begin{lemma}[Detectability of True Edge Differences]
\label{lem:detectability}
Suppose that for every pixel $(i,j) \in \cP$ where $g_0(i,j) \neq g_1(i,j)$, the corresponding error probabilities satisfy $p_{0,0}(i,j) + p_{1,1}(i,j) \neq 1$ (if $g_0(i,j)=0, g_1(i,j)=1$) and $p_{0,1}(i,j) + p_{1,0}(i,j) \neq 1$ (if $g_0(i,j)=1, g_1(i,j)=0$). If $g_0 \neq g_1$, then $\cE(P_0, P_1) > 0$.
\end{lemma}

\begin{proof}
Since $g_0 \neq g_1$, there exists at least one pixel $(x,y) \in \cP$ such that $g_0(x,y) \neq g_1(x,y)$. Without loss of generality, assume $g_0(x,y) = 0$ and $g_1(x,y) = 1$. Then
\[
\E[D_0(x,y)] = p_{0,0}(x,y), \qquad \E[D_1(x,y)] = 1 - p_{1,1}(x,y).
\]
Because $p_{0,0}(x,y) + p_{1,1}(x,y) \neq 1$, we have $\E[D_0(x,y)] \neq \E[D_1(x,y)]$, so the joint distributions $P_0$ and $P_1$ differ in at least one marginal. By the characterization property of energy distance, $P_0 \neq P_1$ implies $\cE(P_0, P_1) > 0$.
\end{proof}

In particular, when the pixelwise error probabilities are invariant across regimes ($p_{0,0}=p_{0,1}=p_0$ and $p_{1,0}=p_{1,1}=p_1$), the condition reduces to $p_0(i,j) + p_1(i,j) \neq 1$. Under stationary in-control noise and fixed pre-processing $\mathcal{C}_{\vartheta}$, any intensity variation that leaves the distribution of $\mathcal{C}_{\vartheta}(Z_t)$ unchanged (in particular, $f^{(0)} = f^{(1)}$) yields $P_0 = P_1$ and hence $\cE(P_0, P_1) = 0$.

\begin{theorem}[Asymptotic Null Distribution]
\label{thm:null_dist}
Let $X_1,\dots,X_{n_1} \overset{\mathrm{iid}}{\sim} P_0$ and $Y_1,\dots,Y_{n_2} \overset{\mathrm{iid}}{\sim} P_1$ in $\R^d$ with $\E\|X_1\|_2^2 < \infty$, and let $N = n_1 + n_2 \to \infty$ such that $n_1/N \to \rho \in (0,1)$. Under $H_0: P_0 = P_1$,
\[
T_{n_1,n_2} \xrightarrow{d} \sum_{r=1}^{\infty} \gamma_r \chi_{1,r}^{2},
\]
where $\{\chi_{1,r}^{2}\}_{r\ge 1}$ are independent $\chi^2(1)$ random variables and $\{\gamma_r\}_{r\ge 1}$ are nonnegative constants satisfying $\sum_{r=1}^{\infty} \gamma_r = \E\|X_1 - X_2\|_2$.
\end{theorem}

\begin{proof}
Under $H_0: P_0 = P_1$, the centered distance kernel of the two-sample $V$-statistic $T_{n_1,n_2}$ is degenerate of order one. Since $\E\|X_1\|_2^2 < \infty$, the result follows from the limit theory for degenerate two-sample $V$-statistics \parencite{NEUHAUS1977,Szekely2004,szekely2013}; non-negativity of $T_{n_1,n_2}$ yields the uncentered weighted $\chi^2(1)$ series with mean $\E\|X_1 - X_2\|_2$. (For binary edge maps in $\{0,1\}^{n^2}$, $\|X_1\|_2 \le n$ is bounded, so all moment conditions hold automatically and only finitely many $\gamma_r$ are nonzero.)
\end{proof}

\subsection{Consistency of the Two-Sample Test}

\begin{theorem}[Consistency]
\label{thm:consistency}
Under the sample-size regime of Theorem~\ref{thm:null_dist}, assume $\E\|X_1\|_2^2 < \infty$ and $\E\|Y_1\|_2^2 < \infty$. Then the two-sample test that rejects $H_0: P_0 = P_1$ for large values of $T_{n_1,n_2}$ is consistent against all fixed alternatives $P_0 \neq P_1$ (in particular, all $g_0 \neq g_1$ under Lemma~\ref{lem:detectability}). Specifically, for any fixed level $\alpha \in (0,1)$,
\[
\lim_{n_1,n_2\to\infty} \Prob_{H_1}\!\left(T_{n_1,n_2} > c_\alpha\right) = 1,
\]
where $c_\alpha$ is the upper $\alpha$-quantile of the limiting null distribution in Theorem~\ref{thm:null_dist}.
\end{theorem}

\begin{proof}
Under $H_1: P_0 \neq P_1$, the population energy distance satisfies $\cE(P_0, P_1) = c > 0$. By the strong law of large numbers for two-sample $V$-statistics, $\cE_{n_1,n_2} \xrightarrow{\mathrm{a.s.}} c > 0$. Since $\frac{n_1 n_2}{n_1 + n_2} \to \infty$ as $N \to \infty$ with $n_1/N \to \rho \in (0,1)$, it follows that $T_{n_1,n_2} = \frac{n_1 n_2}{n_1 + n_2}\cE_{n_1,n_2} \xrightarrow{\mathrm{a.s.}} \infty$. Because $c_\alpha < \infty$ by Theorem~\ref{thm:null_dist}, we obtain $\lim_{n_1,n_2\to\infty} \Prob_{H_1}(T_{n_1,n_2} > c_\alpha) = 1$.
\end{proof}

While Theorems~\ref{thm:null_dist} and \ref{thm:consistency} justify the statistical power of the energy statistic as the sample size grows, online monitoring operates in the finite-$w$ regime with overlapping moving windows. This motivates the parametric bootstrap calibration in Section~\ref{sec:control_limit} rather than relying on the asymptotic $\chi^2$-mixture critical value $c_\alpha$.

\section{Numerical Studies}\label{sec:num}

In this section, we conduct extensive Monte Carlo simulations to evaluate the empirical performance of the proposed Energy Statistic-based image monitoring chart (denoted as \textbf{ES-Chart}) and benchmark it against adapted feature-extraction and nonparametric monitoring schemes from the statistical process control (SPC) literature.

\subsection{Simulation Setup and Synthetic Datasets}\label{sec:sim_setup}

To assess the effectiveness of the proposed monitoring scheme under diverse spatial and intensity transformations, we generate sequences of grayscale images at two spatial resolutions: $128 \times 128$ and $256 \times 256$ pixels. In the uncontaminated state, the background pixel intensity is rendered at $0$, while the foreground shape of interest (an octagram star, a circle, or an amorphous blob) is assigned a baseline intensity value of $192$. For the standalone evaluation of the proposed method, each image is contaminated with independent and identically distributed (i.i.d.) zero-mean Gaussian noise with standard deviation $\sigma \in \{0.05, 0.10\}$ (relative to the normalized $[0, 1]$ intensity scale) and monitored using sliding window lengths of $w \in \{5, 7\}$. Furthermore, to stress-test the proposed method against competing baselines under severe noise contamination, a dedicated comparative study is conducted on the $128 \times 128$ resolution datasets at a high noise level of $\sigma = 0.30$.

For each simulation sequence, the first $14$ frames represent the in-control (IC) state ($\tau = 14$), after which an out-of-control (OC) structural or intensity anomaly is progressively introduced across four benchmark transformation scenarios:

\begin{itemize}
    \item \textbf{Star Line Grow (\texttt{star\_line\_grow}):} Simulates a localized structural defect where a linear crack or scratch incrementally grows within an 8-point star (octagram). The process begins with no visible line during the first $14$ IC frames. Over successive frames, the line extends through five discrete length stages, increasing by $8$ pixels (for $128 \times 128$ images) or $16$ pixels (for $256 \times 256$ images) every $7$ frames, culminating in a continuous line of length $32$ pixels (or $64$ pixels) in the final $14$ frames.
    
    \begin{figure}[H]
        \centering
        \begin{subfigure}{0.19\textwidth}
            \includegraphics[width=\linewidth]{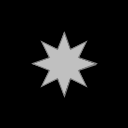}
            \caption{Stage 1 (IC)}
        \end{subfigure}
        \begin{subfigure}{0.19\textwidth}
            \includegraphics[width=\linewidth]{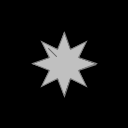}
            \caption{Stage 2}
        \end{subfigure}
        \begin{subfigure}{0.19\textwidth}
            \includegraphics[width=\linewidth]{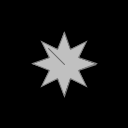}
            \caption{Stage 3}
        \end{subfigure}
        \begin{subfigure}{0.19\textwidth}
            \includegraphics[width=\linewidth]{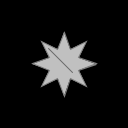}
            \caption{Stage 4}
        \end{subfigure}
        \begin{subfigure}{0.19\textwidth}
            \includegraphics[width=\linewidth]{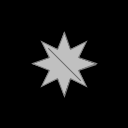}
            \caption{Stage 5}
        \end{subfigure}
        \caption{\texttt{star\_line\_grow} ($128 \times 128$)}
        \label{fig:star_grow}
    \end{figure}
    
    \item \textbf{Star Line Appear (\texttt{star\_line\_appear}):} Displays a series of frames of an octagram where a line of fixed geometry is drawn progressively with increasing opacity, transitioning from near-transparent to fully visible. The first $14$ IC frames feature no line (intensity $192$, matching the octagram). In the three intermediate stages (each spanning $7$ frames), the intensity along the line drops to $170$, $148$, and $126$, respectively, before reaching a highly opaque line of intensity $105$ in the final $14$ frames.
    
    \begin{figure}[H]
        \centering
        \begin{subfigure}{0.19\textwidth}
            \includegraphics[width=\linewidth]{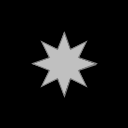}
            \caption{Stage 1 (IC)}
        \end{subfigure}
        \begin{subfigure}{0.19\textwidth}
            \includegraphics[width=\linewidth]{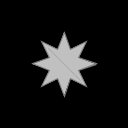}
            \caption{Stage 2}
        \end{subfigure}
        \begin{subfigure}{0.19\textwidth}
            \includegraphics[width=\linewidth]{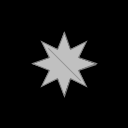}
            \caption{Stage 3}
        \end{subfigure}
        \begin{subfigure}{0.19\textwidth}
            \includegraphics[width=\linewidth]{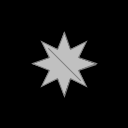}
            \caption{Stage 4}
        \end{subfigure}
        \begin{subfigure}{0.19\textwidth}
            \includegraphics[width=\linewidth]{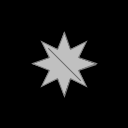}
            \caption{Stage 5}
        \end{subfigure}
        \caption{\texttt{star\_line\_appear} ($128 \times 128$)}
        \label{fig:star_appear}
    \end{figure}
    
    \item \textbf{Circle Shrink (\texttt{circle\_shrink}):} Captures a global geometric deformation where the area of a circle decreases progressively across frames. The initial $14$ IC frames depict a circle with a radius of $32$ pixels (for $128 \times 128$) or $64$ pixels (for $256 \times 256$). The radius then decreases by $4$ pixels (or $8$ pixels) every $7$ frames, reaching a final radius of $16$ pixels (or $32$ pixels) in the last $14$ frames.
    
    \begin{figure}[H]
        \centering
        \begin{subfigure}{0.19\textwidth}
            \includegraphics[width=\linewidth]{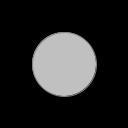}
            \caption{Stage 1 (IC)}
        \end{subfigure}
        \begin{subfigure}{0.19\textwidth}
            \includegraphics[width=\linewidth]{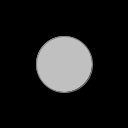}
            \caption{Stage 2}
        \end{subfigure}
        \begin{subfigure}{0.19\textwidth}
            \includegraphics[width=\linewidth]{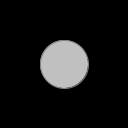}
            \caption{Stage 3}
        \end{subfigure}
        \begin{subfigure}{0.19\textwidth}
            \includegraphics[width=\linewidth]{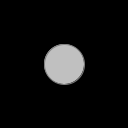}
            \caption{Stage 4}
        \end{subfigure}
        \begin{subfigure}{0.19\textwidth}
            \includegraphics[width=\linewidth]{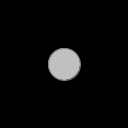}
            \caption{Stage 5}
        \end{subfigure}
        \caption{\texttt{circle\_shrink} ($128 \times 128$)}
        \label{fig:circle_shrink}
    \end{figure}
    
    \item \textbf{Blob Pacman (\texttt{blob\_pacman}):} Models a boundary erosion sequence in which an intact blob gradually morphs into a blob with a wedge-shaped incision (resembling a Pacman profile). The first $14$ IC frames display the intact blob, while the final $14$ frames depict the blob with the full incision, transitioning through three intermediate stages of $7$ frames each.
    
    \begin{figure}[H]
        \centering
        \begin{subfigure}{0.19\textwidth}
            \includegraphics[width=\linewidth]{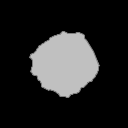}
            \caption{Stage 1 (IC)}
        \end{subfigure}
        \begin{subfigure}{0.19\textwidth}
            \includegraphics[width=\linewidth]{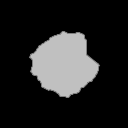}
            \caption{Stage 2}
        \end{subfigure}
        \begin{subfigure}{0.19\textwidth}
            \includegraphics[width=\linewidth]{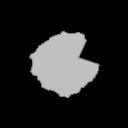}
            \caption{Stage 3}
        \end{subfigure}
        \begin{subfigure}{0.19\textwidth}
            \includegraphics[width=\linewidth]{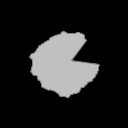}
            \caption{Stage 4}
        \end{subfigure}
        \begin{subfigure}{0.19\textwidth}
            \includegraphics[width=\linewidth]{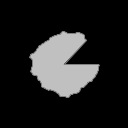}
            \caption{Stage 5}
        \end{subfigure}
        \caption{\texttt{blob\_pacman} ($128 \times 128$)}
        \label{fig:blob_pacman}
    \end{figure}
\end{itemize}

\noindent \textit{(Note: For brevity, only the $128 \times 128$ image figures are shown above. Analogous spatial transformations apply to the $256 \times 256$ datasets.)}

\subsection{Evaluation Metrics and Hyperparameter Tuning}

To evaluate monitoring performance consistently across all experiments, the control limit $h$ is calibrated via the sequential bootstrap bisection procedure of Section~\ref{sec:control_limit} to achieve a nominal in-control Average Run Length of $\mathrm{ARL}_0 \approx 100$. Performance is evaluated over $1,000$ Monte Carlo replications using three metrics:
\begin{itemize}
    \item \textbf{In-Control Average Run Length ($\mathrm{ARL}_0$):} The average number of monitoring windows evaluated before a false alarm occurs under IC conditions.
    \item \textbf{False Detections (FD):} The number of simulation runs (out of $1,000$) in which a false alarm occurs at or before the last in-control frame $\tau = 14$.
    \item \textbf{Mean Detection Delay (MDD):} The average number of frames elapsed after the first out-of-control frame $\tau + 1$ until detection, conditioned on no false alarm prior to the change: $\E[\mathsf{S}(h) - (\tau + 1) \mid \mathsf{S}(h) \ge \tau + 1]$, where $\mathrm{MDD} = 0$ indicates immediate detection upon observing the first out-of-control image.
\end{itemize}

For the proposed \textbf{ES-Chart}, the bootstrap noise scaling hyperparameter $\lambda$ (detailed in Section~\ref{sec:control_limit}) is tuned across each geometric shape, resolution, noise level $\sigma \in \{0.05, 0.10\}$, and window size $w \in \{5, 7\}$. Tables \ref{tab:lambda_256} and \ref{tab:lambda_128} report the calibrated $\lambda$ values for the $256 \times 256$ and $128 \times 128$ resolutions, respectively.

\begin{table}[H]
    \centering
    \caption{Tuned bootstrap hyperparameter $\lambda$ for $256 \times 256$ images.}
    \label{tab:lambda_256}
    \begin{subtable}{0.32\textwidth}
        \centering
        \begin{tabular}{lcc}
            \toprule
            $\sigma \setminus w$ & $5$ & $7$ \\
            \midrule
            $0.05$ & $\sqrt{0.5075}$ & $\sqrt{0.4024}$ \\
            $0.10$ & $\sqrt{0.6775}$ & $\sqrt{0.4125}$ \\
            \bottomrule
        \end{tabular}
        \caption{Star}
    \end{subtable}
    \hfill
    \begin{subtable}{0.32\textwidth}
        \centering
        \begin{tabular}{lcc}
            \toprule
            $\sigma \setminus w$ & $5$ & $7$ \\
            \midrule
            $0.05$ & $\sqrt{1.350}$ & $\sqrt{1.205}$ \\
            $0.10$ & $\sqrt{1.075}$ & $\sqrt{0.825}$ \\
            \bottomrule
        \end{tabular}
        \caption{Circle}
    \end{subtable}
    \hfill
    \begin{subtable}{0.32\textwidth}
        \centering
        \begin{tabular}{lcc}
            \toprule
            $\sigma \setminus w$ & $5$ & $7$ \\
            \midrule
            $0.05$ & $\sqrt{0.795}$ & $\sqrt{0.656}$ \\
            $0.10$ & $\sqrt{0.700}$ & $\sqrt{0.625}$ \\
            \bottomrule
        \end{tabular}
        \caption{Blob}
    \end{subtable}
\end{table}

\begin{table}[H]
    \centering
    \caption{Tuned bootstrap hyperparameter $\lambda$ for $128 \times 128$ images.}
    \label{tab:lambda_128}
    \begin{subtable}{0.32\textwidth}
        \centering
        \begin{tabular}{lcc}
            \toprule
            $\sigma \setminus w$ & $5$ & $7$ \\
            \midrule
            $0.05$ & $\sqrt{0.2825}$ & $\sqrt{0.2175}$ \\
            $0.10$ & $\sqrt{0.1624}$ & $\sqrt{0.2045}$ \\
            \bottomrule
        \end{tabular}
        \caption{Star}
    \end{subtable}
    \hfill
    \begin{subtable}{0.32\textwidth}
        \centering
        \begin{tabular}{lcc}
            \toprule
            $\sigma \setminus w$ & $5$ & $7$ \\
            \midrule
            $0.05$ & $\sqrt{0.503}$ & $\sqrt{0.778}$ \\
            $0.10$ & $\sqrt{0.871}$ & $\sqrt{0.897}$ \\
            \bottomrule
        \end{tabular}
        \caption{Circle}
    \end{subtable}
    \hfill
    \begin{subtable}{0.32\textwidth}
        \centering
        \begin{tabular}{lcc}
            \toprule
            $\sigma \setminus w$ & $5$ & $7$ \\
            \midrule
            $0.05$ & $\sqrt{2.325}$ & $\sqrt{1.775}$ \\
            $0.10$ & $\sqrt{1.381}$ & $\sqrt{1.025}$ \\
            \bottomrule
        \end{tabular}
        \caption{Blob}
    \end{subtable}
\end{table}

\subsection{Performance of the Proposed Method}

Tables \ref{tab:results_256} and \ref{tab:results_128} report the calculated threshold $h$, Mean Detection Delay (MDD), False Detections (FD), and empirical $\mathrm{ARL}_0$ for the proposed \textbf{ES-Chart} on the $256 \times 256$ and $128 \times 128$ datasets under $\sigma \in \{0.05, 0.10\}$ and $w \in \{5, 7\}$.

\begin{table}[H]
    \centering
    \caption{Performance of the proposed ES-Chart on $256 \times 256$ images across noise levels $\sigma \in \{0.05, 0.10\}$ and window sizes $w \in \{5, 7\}$.}
    \label{tab:results_256}
    \begin{tabular}{cc l cccc}
        \toprule
        $\sigma$ & $w$ & Dataset & Threshold ($h$) & Mean Delay (MDD) & False Detections & $\mathrm{ARL}_0$ \\
        \midrule
        \multirow{4}{*}{$0.05$} & \multirow{4}{*}{$5$} 
        & \texttt{star\_line\_grow}   & $8.258$  & $0.871$ & $88$  & $111.860$ \\
        & & \texttt{star\_line\_appear} & $8.258$  & $6.815$ & $88$  & $111.860$ \\
        & & \texttt{circle\_shrink}     & $7.249$  & $0.000$ & $91$  & $106.863$ \\
        & & \texttt{blob\_pacman}       & $6.721$  & $0.183$ & $127$ & $88.705$  \\
        \midrule
        \multirow{4}{*}{$0.10$} & \multirow{4}{*}{$5$} 
        & \texttt{star\_line\_grow}   & $12.075$ & $0.971$ & $94$  & $117.202$ \\
        & & \texttt{star\_line\_appear} & $12.075$ & $6.866$ & $94$  & $117.202$ \\
        & & \texttt{circle\_shrink}     & $9.344$  & $0.000$ & $102$ & $114.505$ \\
        & & \texttt{blob\_pacman}       & $9.046$  & $0.418$ & $123$ & $91.049$  \\
        \midrule
        \multirow{4}{*}{$0.05$} & \multirow{4}{*}{$7$} 
        & \texttt{star\_line\_grow}   & $7.984$  & $0.959$ & $74$  & $107.769$ \\
        & & \texttt{star\_line\_appear} & $7.984$  & $6.883$ & $74$  & $107.769$ \\
        & & \texttt{circle\_shrink}     & $7.498$  & $0.000$ & $90$  & $108.876$ \\
        & & \texttt{blob\_pacman}       & $6.786$  & $0.576$ & $117$ & $93.878$  \\
        \midrule
        \multirow{4}{*}{$0.10$} & \multirow{4}{*}{$7$} 
        & \texttt{star\_line\_grow}   & $11.528$ & $1.179$ & $78$  & $106.543$ \\
        & & \texttt{star\_line\_appear} & $11.528$ & $7.035$ & $78$  & $106.543$ \\
        & & \texttt{circle\_shrink}     & $9.404$  & $0.000$ & $60$  & $123.421$ \\
        & & \texttt{blob\_pacman}       & $9.201$  & $0.688$ & $98$  & $101.245$ \\
        \bottomrule
    \end{tabular}
\end{table}

\begin{table}[H]
    \centering
    \caption{Performance of the proposed ES-Chart on $128 \times 128$ images across noise levels $\sigma \in \{0.05, 0.10\}$ and window sizes $w \in \{5, 7\}$.}
    \label{tab:results_128}
    \begin{tabular}{cc l cccc}
        \toprule
        $\sigma$ & $w$ & Dataset & Threshold ($h$) & Mean Delay (MDD) & False Detections & $\mathrm{ARL}_0$ \\
        \midrule
        \multirow{4}{*}{$0.05$} & \multirow{4}{*}{$5$} 
        & \texttt{star\_line\_grow}   & $5.713$ & $1.264$  & $108$ & $100.739$ \\
        & & \texttt{star\_line\_appear} & $5.713$ & $13.467$ & $108$ & $100.739$ \\
        & & \texttt{circle\_shrink}     & $4.804$ & $0.000$  & $127$ & $93.596$  \\
        & & \texttt{blob\_pacman}       & $7.170$ & $0.858$  & $111$ & $105.310$ \\
        \midrule
        \multirow{4}{*}{$0.10$} & \multirow{4}{*}{$5$} 
        & \texttt{star\_line\_grow}   & $7.722$ & $1.435$  & $97$  & $111.869$ \\
        & & \texttt{star\_line\_appear} & $7.722$ & $13.293$ & $97$  & $111.869$ \\
        & & \texttt{circle\_shrink}     & $6.829$ & $0.000$  & $110$ & $101.626$ \\
        & & \texttt{blob\_pacman}       & $8.248$ & $0.869$  & $124$ & $96.286$  \\
        \midrule
        \multirow{4}{*}{$0.05$} & \multirow{4}{*}{$7$} 
        & \texttt{star\_line\_grow}   & $5.607$ & $1.653$  & $82$  & $108.671$ \\
        & & \texttt{star\_line\_appear} & $5.607$ & $13.726$ & $82$  & $108.671$ \\
        & & \texttt{circle\_shrink}     & $5.278$ & $0.000$  & $90$  & $96.706$  \\
        & & \texttt{blob\_pacman}       & $7.028$ & $0.962$  & $92$  & $103.473$ \\
        \midrule
        \multirow{4}{*}{$0.10$} & \multirow{4}{*}{$7$} 
        & \texttt{star\_line\_grow}   & $7.791$ & $1.825$  & $78$  & $106.449$ \\
        & & \texttt{star\_line\_appear} & $7.791$ & $13.515$ & $78$  & $106.449$ \\
        & & \texttt{circle\_shrink}     & $6.883$ & $0.014$  & $107$ & $95.013$  \\
        & & \texttt{blob\_pacman}       & $8.116$ & $0.983$  & $107$ & $102.136$ \\
        \bottomrule
    \end{tabular}
\end{table}

From Tables \ref{tab:results_256} and \ref{tab:results_128}, the \textbf{ES-Chart} demonstrates near-instantaneous detection across all geometric and boundary transformations: $\mathrm{MDD}$ is virtually $0.000$ for \texttt{circle\_shrink}, $0.183$--$0.983$ for \texttt{blob\_pacman}, and $0.871$--$1.825$ for \texttt{star\_line\_grow}. For \texttt{star\_line\_appear}, the mean delay is approximately $6.8$--$7.0$ frames ($256 \times 256$) and $13.3$--$13.7$ frames ($128 \times 128$), which corresponds directly to the frame at which the decreasing line intensity crosses the lower hysteresis threshold of the Canny edge detector. Furthermore, increasing the window size from $w=5$ to $w=7$ slightly increases MDD by $0.1$--$0.4$ frames due to the inertia of the $w-1$ IC frames in $\mathcal{W}_t$. The lower pre-change FD counts observed for $w=7$ reflect the fact that, with $\tau = 14$ fixed and $\mathrm{ARL}_0$ calibrated to $100$, only $\tau - w + 1 = 8$ in-control windows are evaluated before the change point when $w=7$, compared to $10$ windows when $w=5$.

\subsection{Comparative Study with Competing Methods ($128 \times 128$, $\sigma = 0.30$)}

To rigorously evaluate the robustness of our Canny edge-based dimension reduction and moving-window energy test under severe noise contamination, we benchmark the proposed \textbf{ES-Chart} against three competing schemes adapted from the SPC literature on the $128 \times 128$ resolution datasets at a high noise level of $\sigma = 0.30$ (with sliding window length $w = 5$). 

The first two benchmarks replace our Canny edge extraction and active-pixel projection $M_t$ with frequency-domain wavelet compression while retaining the two-sample Energy Statistic detector, thereby isolating the effectiveness of our spatial pre-processing pipeline. The third benchmark evaluates a supervised learning and nonparametric residual-monitoring paradigm:

\begin{enumerate}
    \item \textbf{1D-Wavelet with Energy Statistic (\textbf{1D-Wav-ES})} \parencite{Koosha2017}: 
    Adapted from the 1D wavelet profile monitoring approach of \textcite{Koosha2017}, each $128 \times 128$ noisy image $Z_t$ is partitioned row-wise into $128$ one-dimensional profiles. A 4-level ($J=4$) 1D Discrete Wavelet Transform (DWT) using the Haar basis is applied to each row to extract $8$ low-frequency approximation coefficients per profile, which are concatenated into a feature vector $\boldsymbol{V}_t \in \mathbb{R}^{1024}$. Whereas \textcite{Koosha2017} monitor these coefficients via a parametric Generalized Likelihood Ratio (GLR) chart, we ensure a direct comparison of feature representations by constructing the reference set $\mathcal{I}^{\mathrm{1D}} = \{\boldsymbol{V}_{-w+1}, \dots, \boldsymbol{V}_0\}$ and sliding test window $\mathcal{W}_t^{\mathrm{1D}} = \{\boldsymbol{V}_{t-w+1}, \dots, \boldsymbol{V}_t\}$, and evaluating the two-sample Energy Statistic $\mathcal{T}_t^{\mathrm{1D}} = (w/2)\widehat{\mathcal{E}}(\mathcal{I}^{\mathrm{1D}}, \mathcal{W}_t^{\mathrm{1D}})$.

    \item \textbf{2D-Wavelet with Energy Statistic (\textbf{2D-Wav-ES})} \parencite{Koosha2022}:
    Following the 2D wavelet framework of \textcite{Koosha2022}, a 2-level ($J=2$) 2D Haar DWT is applied directly to each $128 \times 128$ noisy image $Z_t$, summarizing each $4 \times 4$ pixel block into a single low-frequency coefficient. The resulting $32 \times 32$ approximation sub-band ($A_2$) is vectorized into $\boldsymbol{\Theta}_t \in \mathbb{R}^{1024}$. Rather than applying the Gaussian Likelihood Ratio Test (LRT) of \textcite{Koosha2022}, we evaluate the two-sample Energy Statistic $\mathcal{T}_t^{\mathrm{2D}} = (w/2)\widehat{\mathcal{E}}(\mathcal{I}^{\mathrm{2D}}, \mathcal{W}_t^{\mathrm{2D}})$ between the reference window $\mathcal{I}^{\mathrm{2D}} = \{\boldsymbol{\Theta}_{-w+1}, \dots, \boldsymbol{\Theta}_0\}$ and the moving test window $\mathcal{W}_t^{\mathrm{2D}} = \{\boldsymbol{\Theta}_{t-w+1}, \dots, \boldsymbol{\Theta}_t\}$.

    \item \textbf{Global Residual Anderson--Darling Chart (\textbf{spc4sts-AD})} \parencite{Bui2021}:
    Adapted from the stochastic textured surface (\texttt{spc4sts}) framework of \textcite{Bui2021}, this benchmark models local spatial dependencies in pixel intensities via supervised learning and omits the temporal moving window $\mathcal{W}_t$. Using an in-control reference image, each pixel intensity $y_{0,i}$ is regressed on its causal raster-scan neighborhood $\boldsymbol{y}_0^{(i)}$ via a regression tree to obtain the fitted predictor $\hat{g}_0(\cdot)$ and the empirical in-control residual cumulative distribution function $\hat{\varphi}(\cdot)$. For each online image $Z_t$, prediction residuals $r_{t,i} = y_{t,i} - \hat{g}_0(\boldsymbol{y}_t^{(i)})$ are computed across all $M$ valid interior pixels with complete neighborhoods and ordered as $r_{t,(1)} \leq r_{t,(2)} \leq \dots \leq r_{t,(M)}$. Instead of maximizing over local spatial moving windows, a global one-sample Anderson-Darling (AD) statistic is evaluated across the entire residual image:
    \[A_t^2 = -M - \sum_{k=1}^{M} \frac{2k-1}{M} \left[ \log \hat{\varphi}\left(r_{t,(k)}\right) + \log\left(1 - \hat{\varphi}\left(r_{t,(M+1-k)}\right)\right) \right].\]
\end{enumerate}

\begin{table}[H]
    \centering
    \caption{Comparative performance (Mean Detection Delay, MDD) of the proposed ES-Chart against competing methods on $128 \times 128$ images at $\sigma = 0.30$ and $w = 5$ (calibrated to a nominal $\mathrm{ARL}_0 \approx 100$).}
    \label{tab:competing_comparison_03}
    \begin{tabular}{l cccc}
        \toprule
        & \multicolumn{4}{c}{Mean Detection Delay (MDD)} \\
        \cmidrule(lr){2-5}
        Dataset & \textbf{ES-Chart} & \textbf{1D-Wav-ES} & \textbf{2D-Wav-ES} & \textbf{spc4sts-AD} \\
        \midrule
        \texttt{star\_line\_grow}   & $3.258$ & $14.650$ & $05.011$ & NA \\
        \texttt{star\_line\_appear} & $7.420$ & $18.647$ & $10.518$ & NA \\
        \texttt{circle\_shrink}     & $0.801$ & $00.000$ & $00.000$ & $7.0$ \\
        \texttt{blob\_pacman}       & $2.042$ & $00.111$ & $00.001$ & NA \\
        \bottomrule
    \end{tabular}
\end{table}

Table~\ref{tab:competing_comparison_03} highlights the complementary strengths of competing feature representations under severe noise contamination ($\sigma = 0.30$). On localized structural anomalies (\texttt{star\_line\_grow} and \texttt{star\_line\_appear}), the proposed \textbf{ES-Chart} substantially outperforms both wavelet-based baselines. Since low-frequency wavelet approximation coefficients average intensities over spatial blocks, thin linear cracks and faint boundary transitions are heavily attenuated under high noise. By contrast, Canny edge extraction preserves these fine structural discontinuities. On global area deformations (\texttt{circle\_shrink} and \texttt{blob\_pacman}), where large regions of foreground pixels switch to background intensity, \textbf{1D-Wav-ES} and \textbf{2D-Wav-ES} attain near-zero delays. Finally, \textbf{spc4sts-AD} proves largely ineffective in this setting: it only detects the pronounced global area reduction in \texttt{circle\_shrink}, and even there with a sluggish delay of $7.0$ frames, while behaving essentially randomly and failing to identify the change point on the remaining three scenarios (marked NA). Since \texttt{spc4sts} relies on predicting pixel intensities from local neighborhoods in textured surfaces and evaluates a global single-frame residual statistic without temporal pooling, it is overwhelmed by i.i.d.\ white noise at $\sigma = 0.30$ on piecewise-constant geometries unless the affected pixel area is very large. Overall, \textbf{ES-Chart} offers the most balanced and reliable detection performance across both localized edge defects and global shape deformations.

\subsection{Application to Real Textile Surface Images}\label{sec:real_data}

To further illustrate the practical utility of the proposed \textbf{ES-Chart} in industrial surface inspection, we evaluate its Phase II monitoring performance on real grayscale images of woven textile materials. Monitoring stochastic textured surfaces such as fabric is particularly challenging due to intricate periodic weave patterns, natural background variations, and local discontinuities, which often obscure subtle localized defects from visual inspection.

Following the real-data experimental framework of \textcite{roy2024upper,Bui2018,Bui2021}, we obtain the baseline in-control (IC) textile image at a high spatial resolution of $500 \times 500$ pixels. To simulate an evolving structural defect in a production line, localized ellipsoidal anomalies of progressively increasing spatial extent are induced onto the baseline fabric surface using the \texttt{imposeDefect()} function from the \texttt{spc4sts} package in \textsf{R} \parencite{Bui2021}. 

\begin{figure}[H]
    \centering
    \begin{subfigure}{0.19\textwidth}
        \includegraphics[width=\linewidth]{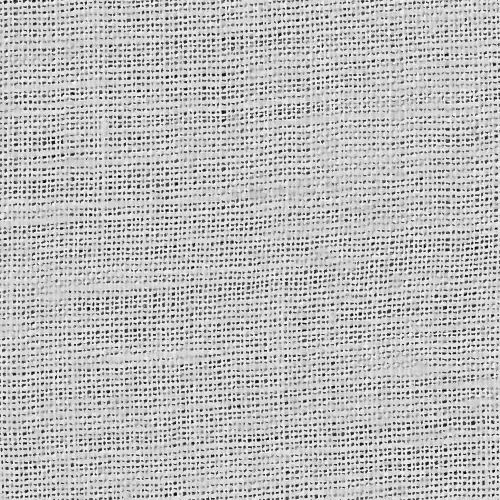}
        \caption{Stage 1 (IC)}
    \end{subfigure}
    \begin{subfigure}{0.19\textwidth}
        \includegraphics[width=\linewidth]{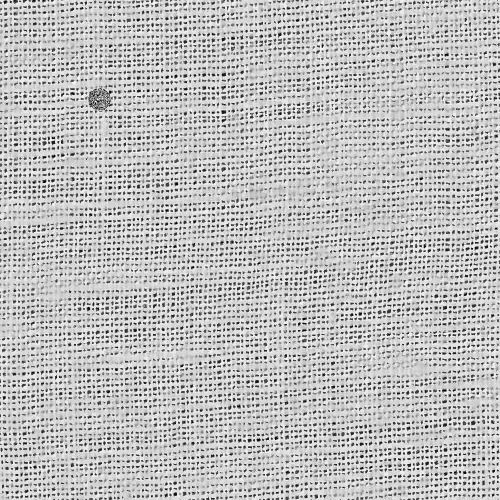}
        \caption{Stage 2 ($11\times 11$)}
    \end{subfigure}
    \begin{subfigure}{0.19\textwidth}
        \includegraphics[width=\linewidth]{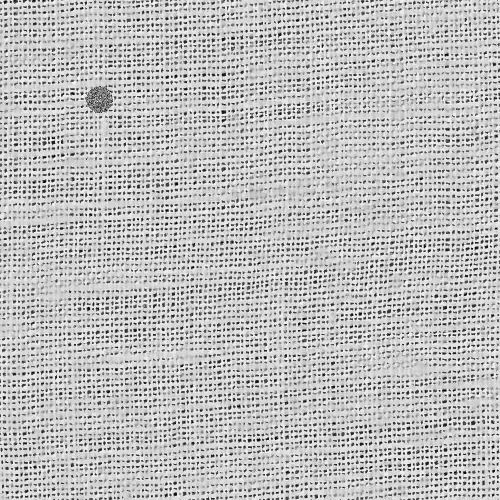}
        \caption{Stage 3 ($13\times 13$)}
    \end{subfigure}
    \begin{subfigure}{0.19\textwidth}
        \includegraphics[width=\linewidth]{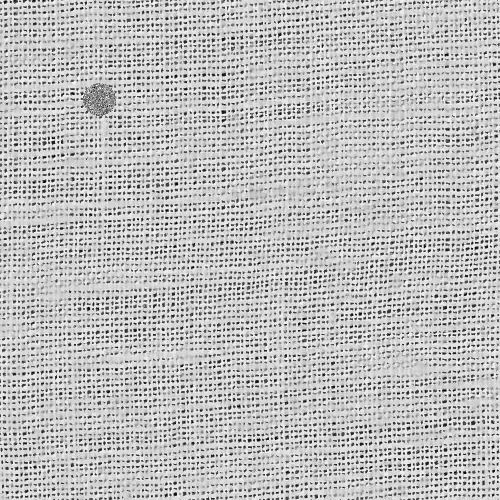}
        \caption{Stage 4 ($16\times 16$)}
    \end{subfigure}
    \begin{subfigure}{0.19\textwidth}
        \includegraphics[width=\linewidth]{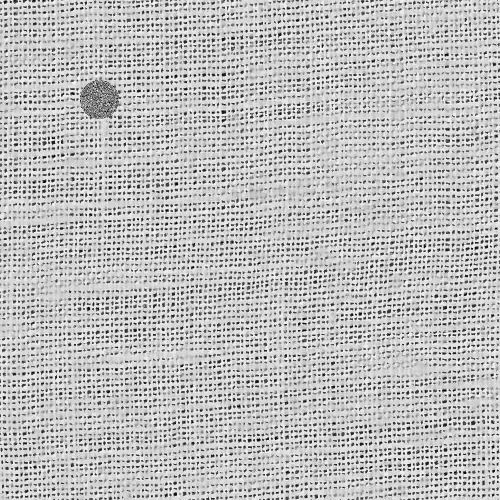}
        \caption{Stage 5 ($19\times 19$)}
    \end{subfigure}
    \caption{Representative frames of the $500 \times 500$ real textile image sequence.}
    \label{fig:textile_stages}
\end{figure}

Specifically, the defect is centered at pixel coordinates $\mathrm{loc} = (100, 100)$ with defect parameters $\mathrm{eps} = 0.05$ and $\mathrm{sigma} = 0.10$, while the semi-axis parameters $(a, b)$ of the ellipsoidal fault region are incrementally expanded over five discrete stages: Stage 1 represents the defect-free IC surface, followed by four out-of-control (OC) stages with $(a, b) = (11, 11)$, $(13, 13)$, $(16, 16)$, and $(19, 19)$, as illustrated in Figure~\ref{fig:textile_stages}.

Similar to the synthetic benchmark sequences in Section~\ref{sec:sim_setup}, the sequential monitoring stream is constructed such that the first $14$ frames represent the in-control process ($\tau = 14$, generated from Stage 1), after which the progressively expanding ellipsoidal defect (Stages 2, 3, and 4, each spanning $7$ frames, followed by Stage 5 for the final $14$ frames) is introduced into the stream, leading to a total sequence of $49$ Phase II frames ($45$ evaluated moving windows for $w = 5$). At each time point $t$, the observed $500 \times 500$ image is contaminated with i.i.d.\ zero-mean Gaussian measurement noise with standard deviation $\sigma = 0.05$.

We monitor the textile image stream using the proposed \textbf{ES-Chart} with a sliding window size of $w = 5$, evaluating empirical performance over $100$ Monte Carlo replications. To accommodate the complex background weave of the fabric during Phase I parametric bootstrap calibration (Section~\ref{sec:control_limit}), the bootstrap noise scaling hyperparameter is tuned to $\lambda = \sqrt{0.03685}$, giving an ARL-calibrated control limit of $h = 127.3933$.

\begin{figure}[H]
    \centering
    \includegraphics[width=0.42\textwidth]{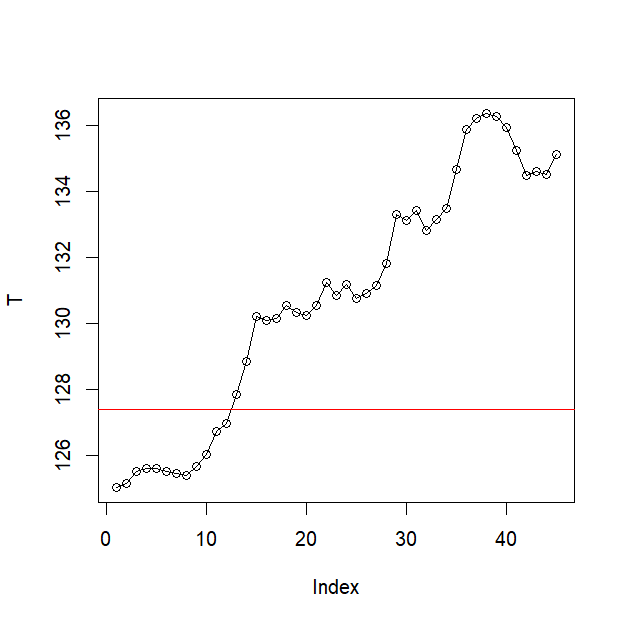}
    \caption{Sample monitoring trajectory of the charting statistic $\mathcal{T}_t$ on the $500 \times 500$ real textile image stream ($w = 5, \sigma = 0.05$). The horizontal red line indicates the calibrated control limit $h = 127.3933$, and the x-axis (Index) represents the evaluated moving-window index $t - w + 1$ (with the first out-of-control frame entering at Index~$= 11$).}
    \label{fig:textile_chart}
\end{figure}

\begin{table}[H]
    \centering
    \caption{Performance of the proposed ES-Chart on the $500 \times 500$ real textile image stream (evaluated over $100$ Monte Carlo replications).}
    \label{tab:textile_results}
    \begin{tabular}{ccccccc}
        \toprule
        $\sigma$ & $w$ & $\lambda$ & Threshold ($h$) & Mean Delay (MDD) & False Detections & $\mathrm{ARL}_0$ \\
        \midrule
        $0.05$ & $5$ & $\sqrt{0.03685}$ & $127.3933$ & $1.950$ & $12$ & $83.040$ \\
        \bottomrule
    \end{tabular}
\end{table}

Table~\ref{tab:textile_results} summarizes the empirical performance of the \textbf{ES-Chart} on the textile dataset, while Figure~\ref{fig:textile_chart} displays a representative Phase II monitoring trajectory of the charting statistic $\mathcal{T}_t$ across the $45$ evaluated windows.

Despite the high dimensionality ($500 \times 500$), the intricate periodic edge structures of the fabric background, and the small spatial footprint of the initial defect relative to the full image frame, the proposed method achieves a low Mean Detection Delay (MDD) of $1.950$ frames. This confirms that the chart reliably triggers an out-of-control alarm well within the very first defect stage, while maintaining a low pre-change false detection count of only $12$ runs (out of $100$ Monte Carlo replications) and an empirical in-control average run length of $\mathrm{ARL}_0 = 83.040$. Overall, these results demonstrate that the boundary-based energy distance statistic scales effectively to high-resolution textured surfaces and remains highly sensitive to subtle localized structural anomalies.

\section{Concluding remarks}\label{sec:conclusion}
In this paper, we have developed a Phase II control chart for image monitoring based on the energy distance between distributions of binary image boundaries. The proposed method follows a fixed reference and moving window framework while introducing an exactly equivalent reduced representation of high-dimensional edge maps to improve computational efficiency. The control limit is calibrated to achieve a prespecified in-control average run length under sequential monitoring. The method is particularly suited to applications in which changes in image boundary structures are of primary interest and geometric alignment is adequately controlled. Potential directions for future research include extending the framework to accommodate spatially dependent image noise and geometric misalignment, as well as developing distance measures that are invariant to nuisance transformations while preserving the mathematical properties required for energy-distance-based monitoring.

\printbibliography

\end{document}